\RequirePackage{fix-cm}
\documentclass[smallcondensed, natbib]{svjour3}       

\journalname{Space Science Reviews}
\smartqed  
\usepackage{graphicx, amsmath}
\newcommand{\nlfop}{\textit{nlf-op}}
\newcommand{\nlfgr}{\textit{nlf-gr}}
\newcommand{\mhs}{\textit{mhs}}
\newcommand{\ffe}{\textit{ffe}}
\newcommand{\mf}{\textit{mf}}
\newcommand{\mhdcese}{\textit{mhd-cese}}
\newcommand{\mhdmas}{\textit{mhd-mas}}
\newcommand{\E}{E}
\newcommand{\Ep}{E_{\rm p}}

\newcommand{\edit}[1]{{#1}}

\usepackage{url}

\begin{document}

\title{Global Non-Potential Magnetic Models of the Solar Corona During the March 2015 Eclipse}

\titlerunning{Global Non-Potential Magnetic Models}        

\author{Anthony R. Yeates \and
Tahar Amari \and
Ioannis Contopoulos \and
Xueshang Feng \and
Duncan H. Mackay \and
Zoran Miki\'c \and
Thomas Wiegelmann \and
Joseph Hutton \and
Christopher A. Lowder \and
Huw Morgan \and
Gordon Petrie \and
Laurel A. Rachmeler \and
Lisa A. Upton \and
Aurelien Canou \and
Pierre Chopin \and
Cooper Downs \and
Miloslav Druckm\"{u}ller \and
Jon A. Linker \and
Daniel B. Seaton \and
Tibor T\"{o}r\"{o}k
}

\authorrunning{A.~R. Yeates et al.} 

\institute{A.~R. Yeates \at
              Department of Mathematical Sciences, Durham University, Science Laboratories, South Road, Durham, DH1 3LE, UK\\
              Tel.: +44-191-3343075\\
              \email{anthony.yeates@durham.ac.uk}   
           \and
           T. Amari \and A. Canou \and P. Chopin \at
              {CNRS, Centre de Physique Th\'eorique de l'Ecole Polytechnique, F-91128 Palaiseau Cedex, France}
           \and
           I. Contopoulos \at
           Research Center for Astronomy and Applied Mathematics, Academy of Athens, Athens 11527, Greece\\
           National Research Nuclear University (MEPhI), Moscow 115409, Russia
           \and
           X. Feng \at
           State Key Laboratory of Space Weather, National Space Science Center, Chinese Academy of Sciences, Beijing 100190, China
           \and
           D.~H. Mackay \at
           School of Mathematics and Statistics, University of St Andrews, North Haugh, St Andrews, Fife, KY16 9SS, UK
           \and
           Z. Miki\'c \and C. Downs \and J.~A. Linker \and T. T\"or\"ok \at
           Predictive Science, Inc., 9990 Mesa Rim Rd., Ste. 170, San Diego, CA 92121-2910, USA
           \and
           T. Wiegelmann \at
           Max-Planck Institut f\"ur Sonnensystemforschung, Justus-von-Liebig-Weg 3, D-37077 G\"ottingen, Germany
           \and
           J. Hutton \and H. Morgan \at
           Institute of Mathematics, Physics \& Computer Sciences, Aberystwyth University, Penglais, Aberystwyth, Ceredigion, SY23 3BZ, UK    
            \and
		C.~A. Lowder \at
		Southwest Research Institute, 1050 Walnut Street, Boulder, CO 80302, USA     
           \and
           G. Petrie \at
           National Solar Observatory, Boulder, CO 80303, USA
           \and
           L.~A. Rachmeler \at
           NASA Marshall Space Flight Center, Huntsville, AL 35811, USA
           \and
           L.~A. Upton \at
           High Altitude Observatory, National Center for Atmospheric Research, 3080 Center Green Dr., Boulder, CO 80301, USA
           \and
           M. Druckm\"{u}ller \at
           Faculty of Mechanical Engineering, Brno University of Technology, 616 69 Brno, Czech Republic
           \and
           D.~B. Seaton \at
Cooperative Institute for Research in Environmental Sciences, University of Colorado, Boulder, CO 80305, USA\\
National Centers for Environmental Information, National Oceanic and Atmospheric Administration, Boulder, CO 80305, USA
              }

\date{Received: date / Accepted: date}

\maketitle

\begin{abstract}

Seven different models are applied to the same problem of simulating the Sun's coronal magnetic field during the solar eclipse on 2015 March 20. All of the models are non-potential, allowing for free magnetic energy, but the associated electric currents are developed in significantly different ways. This is not a direct comparison of the coronal modelling techniques, in that the different models also use different photospheric boundary conditions, reflecting the range of approaches currently used in the community. Despite the significant differences, the results show broad agreement in the overall magnetic topology. Among those models with \edit{significant volume currents in much of the corona}, there is general agreement that the ratio of total to potential magnetic energy should be approximately 1.4. However, there are significant differences in the electric current distributions; while static extrapolations are best able to reproduce active regions, they are unable to recover sheared magnetic fields in filament channels using currently available vector magnetogram data. By contrast, time-evolving simulations can recover the filament channel fields at the expense of not matching the observed vector magnetic fields within active regions. We suggest that, at present, the best approach may be a hybrid model using static extrapolations but with additional energization informed by simplified evolution models. This is demonstrated by one of the models.

\end{abstract}

\section{Introduction}

In recent years, a number of different approaches have been developed for modelling non-potential magnetic fields in the Sun's atmosphere, based on measurements of the magnetic field on the solar surface. Non-potential means that electric currents are allowed to be present within the coronal volume, in contrast to traditional potential-field extrapolations. However, different modellers use not only different approximations for the coronal magnetic field but also different input data and boundary conditions. To date, a direct comparison of these various approaches has been lacking in the literature. With this in mind, we convened a scientific team at the International Space Science Institute in Bern, Switzerland, and we report here on the results.

In this paper, we present a number of non-potential models side-by-side in a way that enables a direct comparison. Since the problem of coronal modelling is very much an active area of research, there are inevitable disagreements between the models. By highlighting the similarities and differences between different modelling approaches, and why they occur, we hope to assist the community in moving toward a single non-potential modelling approach for the corona. We would like to highlight what factors should be taken into account, and what are the biggest uncertainties in existing models.

\begin{figure}
\includegraphics[width=\textwidth]{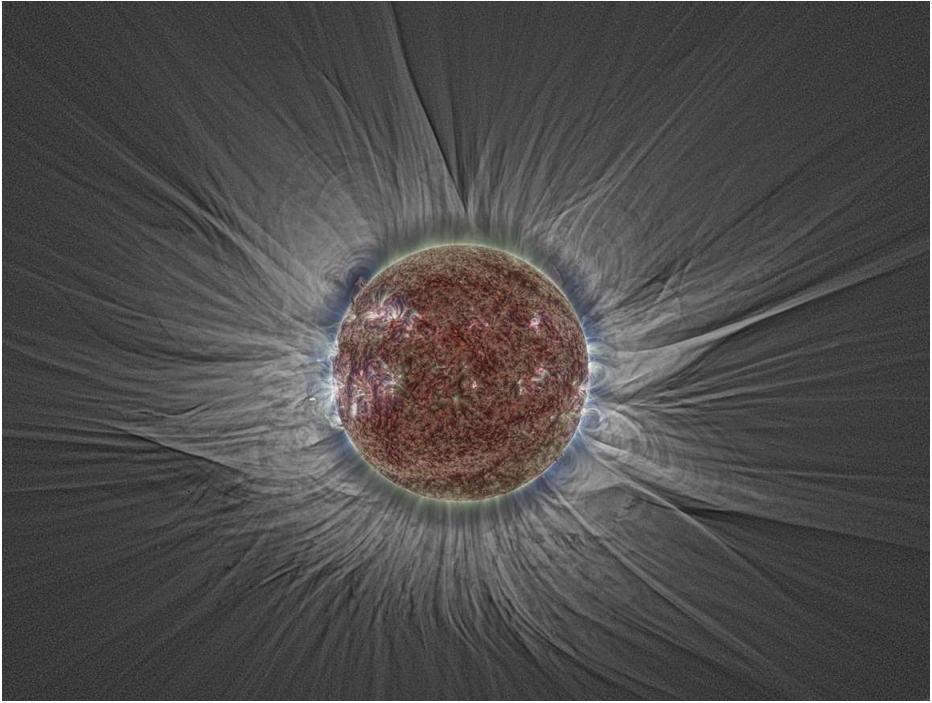}
\caption{An extreme-ultraviolet and white-light composite of the corona on 2015 March 20. The white light image is a combination of 29 exposures made from Longyearbyen, Svalbard, and aligned with sub-pixel precision using the Phase Correlation technique \citep{2009ApJ...706.1605D}. This is overlaid with a combination of 171, 193 and 211 \AA{} (red, green, blue respectively) channels from the AIA instrument on the Solar Dynamics Observatory satellite. Each of the channels were processed using Multiscale Gaussian Normalization \citep[MGN,][]{2014SoPh..289.2945M} prior to combination.}
\label{fig:druckcomp}
\end{figure}

We have fixed a single date and time, coinciding with the total solar eclipse of 2015 March 20. Fixing a single time allows us to compare both static and time-dependent models, and we chose an eclipse date during the Solar Dynamics Observatory (SDO) era so as to maximize the available observations of the real coronal structure. Figure \ref{fig:druckcomp} shows the structure of the observed corona on that day. There are no large active regions on the visible solar disk, although there is some activity on the far side and visible above both East and West limbs. The streamer structure is relatively complex, consistent with the fact that this eclipse occurred  shortly after Solar Maximum. The streamers also show an asymmetry between the north and south poles, consistent with the formation of a polar coronal hole in the south but not in the north \citep{2017SoPh..292...13P}. In terms of non-potential structure, there is a clear polar crown prominence on the north-east limb, with suggestions of a coronal cavity surrounding it. Although this is not a time of particularly high solar activity, the coronal structure is nevertheless more complex than that typically found around Solar Minimum. It is therefore quite challenging to model all of the aspects of this observed corona within the context of a single global magnetic field model.

The compared models are summarized in Section \ref{sec:models}. All cover the full global solar corona, except for the polar regions in some cases. The outer boundaries of the models vary; for the comparisons in this paper we will therefore focus on the region $R_\odot\leq r\leq 2.5R_\odot$ that is covered by all of the models. Grid resolution and boundary conditions were chosen by each individual modeller. In particular, different photospheric boundary conditions were used for each model, and these differences in input data are highlighted in Section \ref{sec:models}. All models used magnetic data from the same SDO/HMI instrument, but different methods were used to reconstruct boundary maps of the full solar surface, leading to quite different photospheric magnetic fields.
Owing to these differences in input data, our study should not be viewed as a direct side-by-side comparison of numerical codes for modelling the coronal magnetic field, such as (for example) the active region exercises by \citet{2008ApJ...675.1637S} and \citet{2015ApJ...811..107D}. Rather, the approach is more qualitative: to compare and contrast the results from different non-potential modelling approaches, where these approaches include the different methods of processing  magnetogram input data that are characteristic of present-day coronal modelling. Through this we can assess the strengths and weaknesses of each approach.

Because we only observe (usually) one side of the Sun, it is difficult to make detailed comparisons with observations using only a single snapshot such as this. Moreover, the lack of direct magnetic measurements above the photosphere is a major driver for the development of these models in the first place. Nevertheless, we are able to compare with indirect observations of the coronal magnetic field, and do so qualitatively in this paper. It is hoped that by combining all available information, we can ultimately build a ``best guess'' picture of the coronal magnetic structure during the eclipse. Following our summary of the models in Section \ref{sec:models}, the resulting coronal magnetic fields are compared in Section \ref{sec:results}, both with each other and with the indirect observations. The overall findings are discussed in Section \ref{sec:discuss}.

\section{Non-Potential Models} \label{sec:models}

\begin{table}
\caption{Summary of the non-potential models presented.}
\label{tab:models}       
\begin{tabular}{p{7cm}ll}
\hline\noalign{\smallskip}
Model & Input Data & Contact \\
\noalign{\smallskip}\hline\noalign{\smallskip}
\nlfop{} -- {Optimization NLFFF} & HMI synoptic ${\bf B}$ & TW\\
 \cite{2007SoPh..240..227W,2014AA...562A.105T} & & \\
\\[-0.1cm]
\nlfgr{} -- {Grad-Rubin NLFFF} & HMI synoptic ${\bf B}$ + HARP  & TA\\
  \cite{2013AA...553A..43A,2014JPhCS.544a2012A} -- XTRAPOLS & &\\
\\[-0.1cm]
\mhs{} -- {Linear MHS} & HMI synoptic $B_r$ & TW \\
  \cite{1986ApJ...306..271B} & & \\
\\[-0.1cm]
\ffe{}  -- {Force-free electrodynamics} & HMI synoptic ${\bf B}$ & IC\\
 \cite{2011SoPh..269..351C,2013SoPh..282..419C} & & \\
\\[-0.1cm]
\mf{} -- {Evolving magnetofrictional} & $B_r$ from HMI-driven AFT$^\ast$ & DHM\\
  \cite{2006ApJ...641..577M, 2014SoPh..289..631Y}  & & \\
\\[-0.1cm]
\mhdcese{} -- {MHD} & HMI synoptic $B_r$ & XF\\
 \cite{2012SoPh..279..207F} -- SIP-AMR-CESE & & \\
\\[-0.1cm]
\mhdmas{} -- {Zero-beta MHD} & HMI synoptic $B_r$ + channels$^\dagger$ & ZM\\
 \cite{1994ApJ...430..898M,1999PhPl....6.2217M} -- MAS & & \\
\noalign{\smallskip}\hline
\end{tabular}
\\[0.2cm]
$\ast$ See Appendix \ref{app:mf}.\\
$\dagger$ Filament channel locations based on \mf{} model (see Appendix \ref{app:mas}).
\end{table}

\begin{table}
\caption{Domains, resolutions and energies$^\ast$ of the non-potential models.}
\label{tab:domains}       
\begin{tabular}{llrrrr}
\hline\noalign{\smallskip}
Model & Boundaries $(\theta, r)$ & Resolution $(r,\theta,\phi)$ & $\E$ [$10^{33}$ ergs] & $\Ep$ [$10^{33}$ ergs] & $\E/\Ep$ \\
\noalign{\smallskip}\hline\noalign{\smallskip}
\nlfop{} & $\pm70^\circ$, $2.56\,R_\odot$ & $180\times 280\times 720$ &3.24 & 2.32 & 1.40\\[0.1cm]
\nlfgr{} &  $\pm90^\circ$, $2.5\,R_\odot$  & $208\times 250\times 500$ & 3.47& 3.42 & 1.01\\[0.1cm]
\mhs{} & $\pm70^\circ$, $2.56\,R_\odot$  & $180\times 280\times 720$ & 3.30 & 2.32 & 1.42\\[0.1cm]
\ffe{} &  $\pm90^\circ$, $5.8 R_\odot$ & $125\times 225\times 450$ & 3.78 & 3.50 & 1.08\\[0.1cm]
\mf{} & $\pm89^\circ$, $2.5\,R_\odot$  & $56\times 180\times 360$ & 2.63& 1.79 & 1.47\\[0.1cm]
\mhdcese{} & $\pm90^\circ$, $30.0\,R_\odot$  & $61^\dagger\times 92\times 182$ & 1.99& 1.65 & 1.21 \\[0.1cm]
\mhdmas{} & $\pm90^\circ$, $2.5\,R_\odot$ & $182\times 200 \times 560$ & 2.52& 1.72 & 1.47\\
\noalign{\smallskip}\hline
\end{tabular}
\\[0.2cm]
$\ast$ Energies are computed between $r=R_\odot$ and $r=2.5\,R_\odot$.\\
$\dagger$ The resolution of 61 points in $r$ refers to the region $r\le 2.5\,R_\odot$ only.
\end{table}

The models presented in this paper are summarized in Table \ref{tab:models}, and their computational domains are summarized in Table \ref{tab:domains}. The latter table also shows magnetic energies, which will be discussed in Section \ref{sec:results}. For purposes of illustration we have selected a single representative model of each type, although additional runs were carried out in many cases. These models are still under development, and not yet freely available, but readers interested in their use are encouraged to contact the individual modellers, as listed in Table \ref{tab:models}.

Space precludes a detailed description of each model, so the following subsections concentrate on their similarities and differences. For a review of the various non-potential modelling approaches, see \citet{2012LRSP....9....6M} or \citet{2017SSRv..210..249W}, or follow the references given in Table \ref{tab:models}. Some additional details are given for the \mhdmas{} and \mf{} models in Appendices \ref{app:mas} and \ref{app:mf} respectively, since these differ somewhat in their setup from previously published simulations.

We stress that, although we consider only a single model of each type, it would be possible to ``mix and match'' aspects of different models in future. For example, the techniques used to energize filament channels in the \mhdmas{} model could equally be applied to any of the other static models. Or thermodynamics effects could be included in the \mhdmas{} model. Thus our comparison should be treated only as a representative selection, aiming to illustrate the consequences of including different model features.

\subsection{Modelling approaches}

In the low corona, on a global scale, magnetic forces dominate over gravity and pressure gradients. Thus the large-scale magnetic field configuration is expected to be approximately force-free at the heights $r< 2.5\,R_\odot$ of interest here \citep{2012LRSP....9....5W}. This means that the Lorentz force vanishes and the magnetic field ${\bf B}$ satisfies
\begin{equation}
{\bf j}\times{\bf B} = 0, \qquad \nabla\cdot{\bf B}=0,
\label{eqn:ff}
\end{equation}
where ${\bf j}=c\nabla\times{\bf B}/(4\pi)$ is the electric current density. In a non-potential magnetic field with free energy, we must have ${\bf j}\neq 0$ at least in some regions of the corona. All of the models in Table \ref{tab:models} aim to approximate force-free equilibria of the form \eqref{eqn:ff}, except for \mhs{} and \mhdcese{} where plasma pressure and gravity also influence the equilibria. Namely, the \mhs{} model uses a special class of magnetohydrostatic equilibria where the electric current density flows on spherical shells. Here we set the parameter $a=1.0$ \citep[for more details and the definition of $a$, see][]{1986ApJ...306..271B}. In this model, the radial magnetic field $B_r$ on $r=R_\odot$ is used as a boundary condition for spherical harmonic decomposition. The resulting magnetic field contains a non-vanishing Lorentz force, which is compensated by pressure gradients and gravity. The density and plasma pressure distribution are computed as a superposition of these terms and a stratified solar atmosphere model. The \mhdcese{} model solves the full magnetohydrodynamic (MHD) equations including gravity, centrifugal and Coriolis terms, as well as a volumetric heating term, relaxing to a static equilibrium that extends out to $r=30\,R_\odot$, albeit at a lower resolution than the other models. All of the other models have an outer boundary at or near $r=2.5\,R_\odot$, except for \ffe{} where the computational domain extends to $5.8\,R_\odot$, although a perfectly-matched (absorbing) layer is implemented above $2.5\,R_\odot$.

Whilst the \nlfop{}, \nlfgr{}, \ffe{}, \mf{} and \mhdmas{} models all use some form of iteration toward ${\bf j}\times{\bf B}=0$ -- constrained by the imposed photospheric boundary conditions -- they differ significantly in how the electric current structures in the final magnetic field are built up. In the first three models, the electric current distribution is fixed by the photospheric boundary conditions. By contrast, in the \mf{} model the coronal magnetic field is driven by large-scale horizontal flows in the photosphere, at the same time as relaxing the coronal magnetic field toward ${\bf j}\times{\bf B}=0$. This leads to an evolving photospheric $B_r$ distribution. The advantage of this approach is that it can build up large-scale current structures in a manner mimicking that in the real Sun. But the disadvantage is that a whole time series of input data extending over several months is required on the photospheric surface, both for $B_r$ and the horizontal flows. Moreover, the final $B_r$ distribution in the \mf{} model matches that of the original magnetogram only on a coarse scale. The \mhdmas{} model is in some sense a hybrid: although it uses a single observed map of $B_r$ on the solar surface, additional electric fields are imposed to drive the formation of filament channels at specified locations in the corona. The locations of these channels, and in particular their chirality (direction of the transverse magnetic field) were guided by the results of the \mf{} model and by EUV observations (see Appendix \ref{app:mas}).

\subsection{Photospheric boundary conditions}

The different modelling techniques each require particular input data at the solar photosphere. All of the models except \mf{} use only a single magnetic map covering the full photospheric surface, notionally at the time of the eclipse.  
Vector magnetic data with all three components of ${\bf B}$ are required for \nlfop{}, \nlfgr{} and \ffe{}, whereas only the radial component $B_r$ is utilized by all of the other models.

Since the observed vector magnetograms are taken at photospheric heights where the magnetic field is not generally force free, these are generally not consistent with \eqref{eqn:ff}. The three models that use vector magnetic input deal with this in different ways. The \nlfgr{} model uses the Grad-Rubin scheme. By imposing the force-free function $\alpha=j_r/B_r$ in only one polarity (here, $B_r>0$), as well as $B_r$ everywhere, it matches the transverse components of ${\bf B}$ in locations of this polarity. Some cut-offs on the $B_r$ and on the transverse components of ${\bf B}$ are applied so as to generate regular values of $\alpha$, and in particular it is assumed that $j_r=0$ in regions of weak $B_r$. The \nlfop{} method uses an optimization that penalizes differences in the simulated and observed vector ${\bf B}$ over the whole map, but allows differences, particularly in pixels with smaller observed transverse field. The \ffe{} method imposes the vector ${\bf B}$ over the whole map, although this leads to a boundary layer just above the photosphere where ${\bf B}$ becomes more force free.

The \mhs{}, \mf{}, \mhdcese{} and \mhdmas{} models require only the observed radial magnetic field $B_r$ on the photospheric boundary rather than the full vector ${\bf B}$, although as mentioned above the \mf{} model requires a whole time sequence of $B_r$ measurements. In the Bogdan-Low solution used by the \mhs{} model, the transverse components of ${\bf B}$ on the boundary are uniquely determined by the form of the solution. However, in the methods used by \mf{}, \mhdcese{} and \mhdmas{}, a transverse electric field is required to specify a unique evolution from the initial potential field.

This transverse electric field is set either by boundary flows or through ad-hoc relations. In \mhdcese{}, the transverse electric field is simply set to zero as the corona is evolved toward equilibrium (the initial plasma parameters are taken from Parker's hydrodynamic isothermal solar wind solution). In \mhdmas{}, the transverse electric field is set to zero except where it is imposed to create magnetic flux ropes in filament channels (see Appendix \ref{app:mas}). The \mf{} model is rather different since  transverse electric fields are imposed continually throughout the 200-day evolution, so as to evolve the photospheric $B_r$ distribution as well as the coronal magnetic field. This is described in more detail in Appendix \ref{app:mf}.

\subsection{Magnetogram input data} \label{sec:mag}

All models used photospheric magnetogram data from the HMI instrument on Solar Dynamics Observatory \citep{2012SoPh..275..229S}. However, it is important to note that these data were processed in different ways by each modeller, leading to different distributions of $B_r$ on $r=R_\odot$. Rather than standardize this processing and do a direct model comparison, we have decided to highlight the fact that the input data themselves represent a significant difference between present models of the coronal magnetic field. The differences are summarized in this section.

\begin{figure}
\centering
\includegraphics[width=\textwidth]{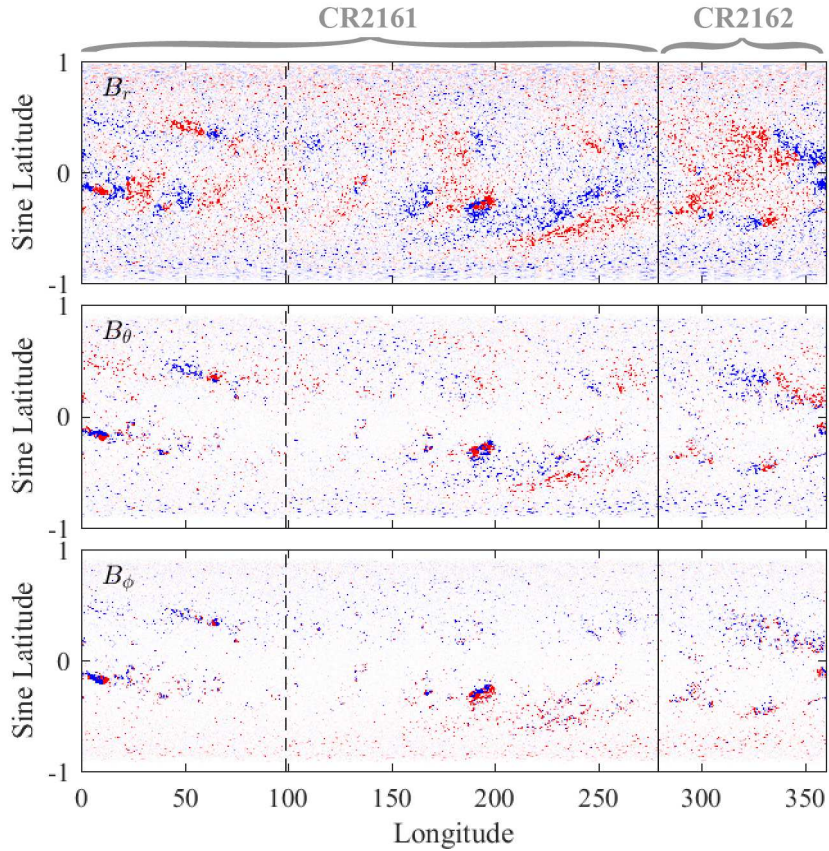}
\caption{The $B_r$, $B_\theta$, and $B_\phi$ components of the \edit{custom} SDO/HMI vector map, with color tables saturated at $\pm 50\,\mathrm{G}$ (red positive, blue negative). The horizontal axis shows Carrington longitude and the vertical dashed line marks the location of central meridian at the time of eclipse ($99^\circ$). The vertical solid line marks the time discontinuity between data from the synoptic maps for CR2161 (to the left) and CR2162 (to the right). \edit{This is at $279^\circ$ rather than the typical $360^\circ$.}}
\label{fig:petrie}
\end{figure}

Several of the models used a \edit{custom} SDO/HMI vector map that was created specially for this project, shown in Figure \ref{fig:petrie}. Here data from the successive Carrington rotations CR2161 and CR2162 were combined so as to extract a synoptic map centered on the eclipse Central Meridian (Carrington longitude $\approx 99^\circ$ during CR2161), with the discontinuity located $180^\circ$ out of phase from this (at Carrington longitude $279^\circ$). The data were processed using the standard HMI pipeline \citep{2014SoPh..289.3483H}, but with azimuth-angle disambiguated full-disk images. Strong fields ($|{\bf B}|\geq 50\,\mathrm{G}$) were left unsmoothed, but boxcar smoothing was applied to low-latitude quiet-Sun fields ($|{\bf B}|< 50\,\mathrm{G}$), in order to improve the signal-to-noise ratio.
High latitude fields (poleward of $\pm 70^\circ$ latitude) were filled using a combination of smoothed vector $B_r$ and radially-corrected line-of-sight field.

For the \ffe{} model, this \edit{custom} vector map (with dimensions $1440\times 3600$) was simply \edit{boxcar} binned to a lower resolution. To generate the initial potential fields for the \nlfop{} and \mhs{} models, the map was initially smoothed by retaining only spherical harmonics of degree $\ell \leq 25$. The full map was then used in the ensuing iteration for \nlfop{}, while the \mhs{} input map remains smoothed. For \nlfop{}, note that no pre-processing of the vector data \citep[as described by][]{2009A&A...508..421T} was applied.

The \nlfgr{} model took the basic map as a starting point but made two modifications. Firstly, $B_r$ on the visible disk (centred at Carrington longitude $99^\circ$) was replaced by an HMI longitudinal \edit{$720$-second} full-disk magnetogram from 09:47:58UT on 2015 March 20. Secondly, six active regions on the visible disk were replaced by HMI vector \edit{Space-Weather HMI Active Region Patches \citep[HARPs;][]{2014SoPh..289.3549B}}. Specifically, patches number 5337, 5342, 5345, 5347, 5348, and 5350, again observed at 09:47:58UT. These additions were made on a higher resolution grid ($3804\times 5784$), which was subsequently interpolated to a final mesh of $250\times 500$ for the computation. This final mesh is non-uniform with more grid points within active regions.

The \mhdcese{} model used \edit{a standard} SDO/HMI synoptic magnetogram for CR2161, smoothed with a simple Gaussian kernel (as a function of heliocentric angle). The \mhdmas{} model used a $B_r$ map constructed from HMI data in a similar way to the \edit{custom} map, combining synoptic maps from CR2161 (from $0^\circ$ to $279^\circ$) and CR2162 (from $279^\circ$ to $360^\circ$). The high-latitude fields were filled in from previous Carrington rotations when the poles were visible, and the maps were suitably smoothed for the MHD calculation. The \mf{} model did not use HMI data directly, being driven instead by a flux transport model coupled with the emergence of bipolar magnetic regions starting from 2014 September 1 -- this is described in Appendix \ref{app:mf}.

It is important to note that the models using synoptic input data from both CR2161 and CR2162 do not represent what could have been achieved in an advance prediction of the eclipse, since the maps incorporate magnetogram data taken after the eclipse time. The exceptions to this are \mf{} and \mhdcese{}, which use only information from the visible face of the Sun up to but not after the eclipse.

\begin{figure*}
\includegraphics[width=\textwidth]{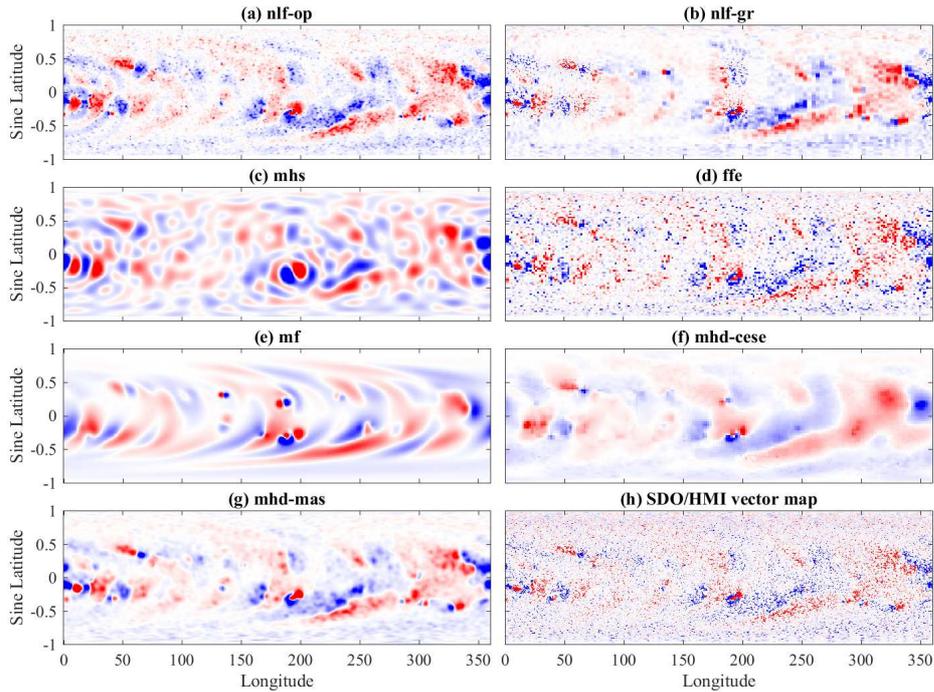}
\caption{Radial magnetic field $B_r(R_\odot,\theta,\phi)$ in each model, over the full solar surface. Here the horizontal axis shows Carrington longitude, with Central Meridian at eclipse time located at $99^\circ$. The color tables are saturated at $\pm 50\,\mathrm{G}$ (red positive, blue negative). Panel (h) repeats $B_r$ from the SDO/HMI map in Figure \ref{fig:petrie}, for ease of comparison.}
\label{fig:br1}
\end{figure*}

Figure \ref{fig:br1} shows the distribution of radial magnetic field at the photosphere $r=R_\odot$ in each model (at the closest simulated time to the eclipse for the \mf{} model). It is clear that there are differences between these input data for the different models. Most obvious are the variations in resolution and in the amount of smoothing applied to the data, as described above. But important also are the timing variations, namely how recently different regions of the maps have been updated. There are two notable differences that affect the model results presented below. Firstly, a small active region (NOAA 12304) is located at approximately $140^\circ$ Carrington longitude, in the Northern hemisphere. This region is visible in the EUV eclipse image (Figure \ref{fig:druckcomp}), in the Northern hemisphere between disk center and the West limb. However, it is missing from the HMI vector synoptic map in Figure \ref{fig:petrie} because it emerged after that Carrington longitude had rotated past Central Meridian. Accordingly, it is omitted from models \nlfop{}, \mhs{}, \ffe{}, \mhdcese{} and \mhdmas{}. It is included, however, in \nlfgr{} thanks to the replacement of the visible disk by an HMI full-disk magnetogram taken just prior to the eclipse. The region is also included in the \mf{} model owing to the method of bipolar magnetic region (BMR) insertion (Appendix \ref{app:mf}).

The second difference is potentially more significant: all models except for \mf{} and \mhdcese{} include the activity complex centred around $10^\circ$ Carrington longitude in the Southern hemisphere, and appearing at the East limb in Figure \ref{fig:druckcomp}. In the \mf{} model the corresponding BMRs have not yet been assimilated at the time of eclipse, since they were not yet fully visible at that time and their properties could not be determined. Similarly they do not appear in \mhdcese{} since the corresponding longitudes were taken from earlier in CR2161, rather than CR2162 as in the other models. Such differences are typical of coronal magnetic field modelling based on synoptic magnetic field observations, but must be taken into account when assessing our results, which we consider next. We reiterate that the inclusion of this activity complex in the other models was possible only by using observations taken after the eclipse; if the comparison had been carried out in real time then this correction would not have been possible.

\section{Model Comparison} \label{sec:results}

A number of aspects of the models are compared in the following subsections, with the aim of highlighting the similarities and differences between them. Since many of the models do not include realistic thermodynamics (or indeed any plasma at all), we limit our comparison to the magnetic field, rather than plasma density or temperature. Some of the models, such as the \mhdmas{} model, have already been developed extensively to describe the flow of energy in the corona, as well as the acceleration of the solar wind, and can deduce the coronal temperature and density. In fact, we also studied this eclipse using this more sophisticated version of the \mhdmas{} model, but we do not describe those results here since they exceed the scope of the comparisons presented in this paper. Ground truth observations of the real corona are included, although these can provide only indirect, qualitative information about the coronal magnetic field. 

\subsection{Magnetic Flux and Energy} \label{sec:flux}

\begin{figure}
\includegraphics[width=\textwidth]{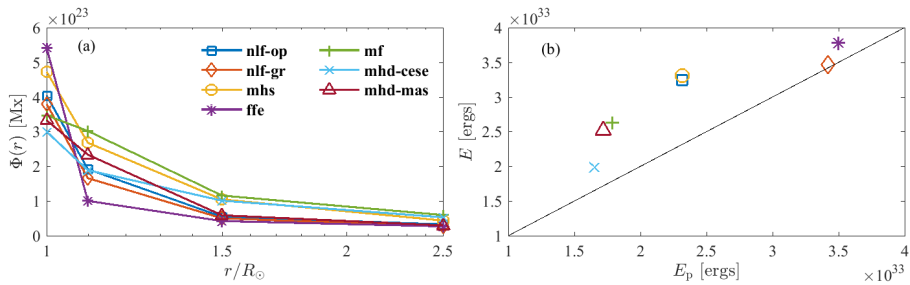}
\caption{Magnetic flux (a) and energy (b) of the models. Shown are (a) unsigned radial magnetic flux $\Phi(r)$ as a function of height, and (b) total magnetic energy $\E$ versus potential magnetic energy $\Ep$. The solid line indicates $\E=\Ep$.}
\label{fig:flux}
\end{figure}

We begin with some overall diagnostics. Firstly, Figure \ref{fig:flux}(a) shows the total unsigned magnetic flux,
\begin{equation}
\Phi(r) = \int_0^{2\pi}\int_{\theta_{\rm min}}^{\theta_{\rm max}}\big|B_r(r,\theta,\phi)\big|r^2\sin\theta\,\mathrm{d}\theta\,\mathrm{d}\phi,
\label{eqn:phi}
\end{equation}
for each model at a sequence of heights in the corona. At $r=R_\odot$ there are significant differences in the photospheric flux arising from the difference in resolution between the models (these different resolutions are stated in Table \ref{tab:domains} and evident in Figure \ref{fig:br1}). The model photospheric fluxes span the range $3-5.5 \times 10^{23}\,\mathrm{Mx}$, from \mhdcese{} with the lowest flux to \ffe{} with the highest. For comparison, note that the SDO/HMI synoptic map in Figure \ref{fig:petrie} has a flux $\Phi(R_\odot)=5.8\times 10^{23}\,\mathrm{Mx}$ at its original resolution. The difference in resolution between the models becomes insignificant above  $1.5\,R_\odot$, where the behaviour is dominated by low-order spherical harmonics. Accordingly, most of the models predict a similar open magnetic flux of around $3\times 10^{22}\,\mathrm{Mx}$. The exceptions are \mhs{}, \mf{} and \mhdcese{}, where the open flux is inflated due to the presence of significant volumetric currents in the corona -- up to $6\times 10^{22}\,\mathrm{Mx}$ in the case of \mf{}. The \mf{} model includes the ejection of magnetic flux ropes which also increases the open flux. \edit{These open flux values, and} the open magnetic field distribution, will be further discussed in Section \ref{sec:open}.

Figure \ref{fig:flux}(b) shows the total magnetic energy of each model between $r=R_\odot$ and $r=2.5R_\odot$, given by
\begin{equation}
\E = \int_{R_\odot}^{2.5R_\odot}\int_0^{2\pi}\int_{\theta_{\rm min}}^{\theta_{\rm max}}\frac{|{\bf B}(r,\theta,\phi)|^2}{8\pi}r^2\sin\theta\,\mathrm{d}\theta\,\mathrm{d}\phi\,\mathrm{d}r,
\end{equation}
and plotted against the corresponding energy $\Ep$ for a potential field source surface extrapolation with source surface at $r=2.5R_\odot$ and the same $B_r(R_\odot,\theta,\phi)$ for each particular model. These energies are also listed in Table  \ref{tab:domains}. There are considerable variations in both $\E$ and $\Ep$, which depend to some extent on the model resolution. The distance of each symbol above the solid line indicates the excess of non-potential energy $\E$ above $\Ep$, or the relative free energy. This varies from very little free energy for \nlfgr{} to around 40\% for \nlfop{}, \mhs{}, \mf{} and \mhdmas{}. Whilst the latter models all produce a similar percentage of free energy, the absolute values differ significantly, and we will see in Section \ref{sec:j} that the distributions of electric current within the volume also differ significantly. In the case of the \mhs{} model, the free energy depends on the choice of the $a$ parameter; for example, taking $a=2$ would lead to $\E=1.73\,\Ep$ rather than $1.42\,\Ep$ for $a=1$.  The particularly low free energy for \nlfgr{} arises because the electric currents are strongly concentrated within active regions, with the magnetic field being close to potential throughout most of the volume.

\subsection{Electric Currents} \label{sec:j}

\begin{figure}
\includegraphics[width=0.7\textwidth]{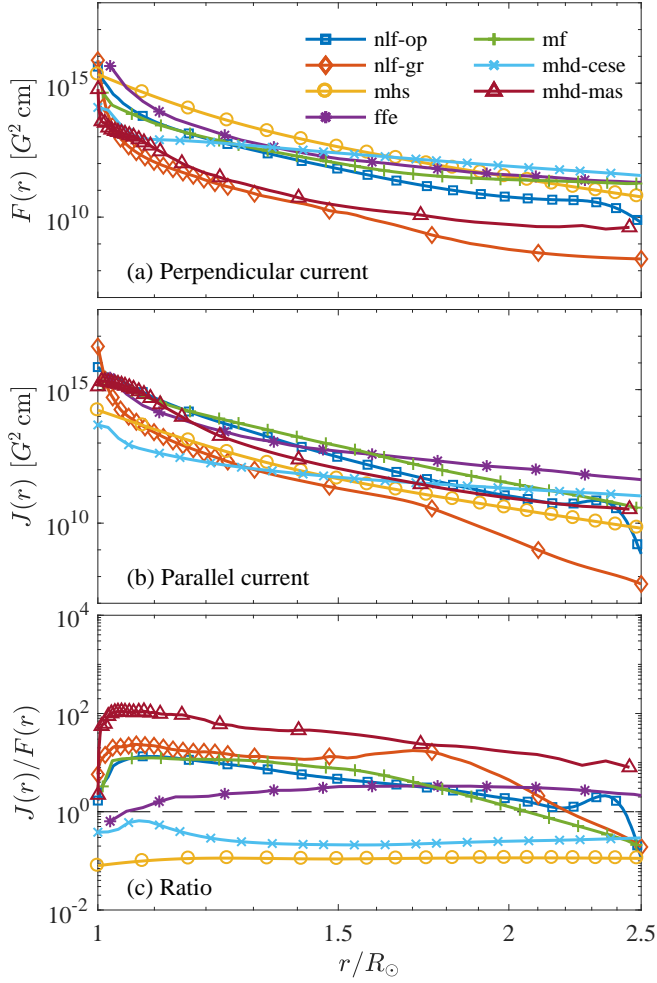}
\caption{Measures of the (a) perpendicular and (b) parallel currents for each model, as functions of height. \edit{Panel (c) shows the ratio of parallel to perpendicular measures.}}
\label{fig:jf}
\end{figure}

\begin{figure}
\includegraphics[width=\textwidth]{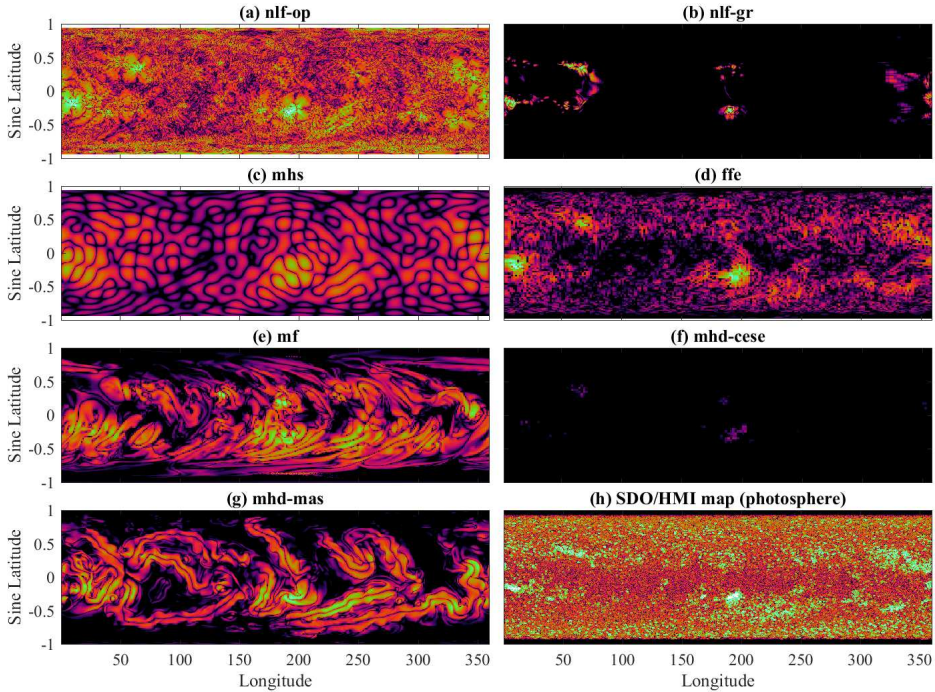}
\caption{Magnitude of the vertical electric current density, $\log_{10}|j_r|$, at $r=1.02R_\odot$ in each model, over the whole spherical surface. The color tables are saturated at $10^{-10}\,\mathrm{G}\,\mathrm{cm}^{-1}$ (black) and $10^{-6.5}\,\mathrm{G}\,\mathrm{cm}^{-1}$ (white). Panel (h) shows the same quantity computed at $r=R_\odot$ directly from the SDO/HMI synoptic map in Figure \ref{fig:petrie} (boxcar smoothed to reduce noise). The same color scale is used.}
\label{fig:jr}
\end{figure}

Figures \ref{fig:jf}(a) and (b) show measures of the average perpendicular and parallel electric currents in each model, as a function of radius. Specifically we plot the quantities
\begin{equation}
F(r) = \int_0^{2\pi}\int_{\theta_{\rm min}}^{\theta_{\rm max}}\big|(\nabla\times{\bf B})\times{\bf B}\big|r^2\sin\theta\,\mathrm{d}\theta\,\mathrm{d}\phi
\end{equation}
and
\begin{equation}
J(r) = \int_0^{2\pi}\int_{\theta_{\rm min}}^{\theta_{\rm max}}\big|(\nabla\times{\bf B})\cdot{\bf B}\big|r^2\sin\theta\,\mathrm{d}\theta\,\mathrm{d}\phi.
\end{equation}
\edit{The perpendicular measure $F(r)$ is essentially a measure of the Lorentz force ${\bf j}\times{\bf B}$. Figure \ref{fig:jf}(c) shows the ratio $J(r)/F(r)$, which is a measure of  ``force-freeness'' for each model.}

\edit{Below about $r=1.5\,R_\odot$, the ratio $J/F$ divides the models into three broad classes: those models that are relatively force-free with $J/F\gg 1$ (\mhdmas{}, \nlfgr{}, \nlfop{} and \mf{}), those that are not force-free, with $J/F < 1$ (\mhdcese{} and \mhs{}), and the \ffe{} model which is not force-free at the photosphere but becomes rather more so above about $1.1\,R_\odot$. These differences arise from the physics of the models: the \mhdcese{} and \mhs{} models include non-magnetic terms in their force balances, while the \ffe{} model matches to a boundary condition at $r=R_\odot$ that does not satisfy ${\bf j}\times{\bf B}=0$. The model with highest ratio $J/F$ is \mhdmas{}, owing to the relatively smooth boundary data and level of numerical relaxation applied.}

\edit{The actual amount of current near the photosphere varies between models, as seen by $J(r)$ in Figure \ref{fig:jr}(b). This is highest for \nlfgr{}, because it includes the most fine-scale structure in the HARP patches. On the other hand, $J(r)$ falls off rapidly with height in this model, consistent with the field being closest to potential overall (Table \ref{tab:domains} and Figure \ref{fig:flux}b). By contrast, the \mhs{} model has relatively low $J(r)$ near the photosphere, owing to its lower input resolution, and the lowest $J(r)$ is found for the \mhdcese{} model, which lacks a mechanism for energizing the magnetic field there; its free magnetic energy is located at larger radii where the non-magnetic terms become important.}

\edit{Both the \mf{} and \mhdmas{} models have lower $J(r)$ very close to the photosphere, because they do not resolve such fine-scale structures as \nlfgr{} and \nlfop{}. However, these two models have larger current in the low corona ($1.1$ to $1.2\,R_\odot$), arising from the incorporation of low-lying filament channels. These form self-consistently in the \mf{} model, while their locations (for imposed currents) were chosen by design in the \mhdmas{} model. Between about $1.2$ and $1.5\,R_\odot$, the \mf{} model has the largest $J(r)$, owing to the significant volume currents that have been ejected over time by flux emergence and surface motions. As in the \nlfop{} model, which has almost as high $J(r)$, these currents are not limited to active regions, and reach greater heights in the corona.} 

\edit{Above about $1.7\,R_\odot$, the \ffe{} model has the largest $J(r)$, although this likely results from incomplete relaxation to equilibrium in the model and deserves further investigation. The \mhdcese{} model also has a more gradual fall-off of $J(r)$ with height than many of the other models, again reflecting the influence of non force-free effects at larger heights in the model as the magnetic field is opened out by the solar wind. Similarly, the \mf{} model becomes non force-free ($J/F<1$) above about $2\,R_\odot$,  caused by the imposed outflow at the upper boundary. This outflow is used to simulate the effect of the solar wind \citep{2006ApJ...641..577M}, so that the magnetic field is no longer force-free above $2\,R_\odot$ but is in a steady-state balance between the outflow and the Lorentz force.}

To illustrate the spatial distributions of currents, Figure \ref{fig:jr} shows the magnitude of vertical current density $j_r(1.02\,R_\odot,\theta,\phi)$ in each model, with the same logarithmic color scale used for all plots. For comparison, panel (h) shows the magnitude of $j_r(R_\odot,\theta,\phi)$ computed directly from the SDO/HMI vector synoptic map (Figure \ref{fig:petrie}).

In all models, the strongest currents are located in the active regions, as might be expected since the magnetic field is strongest there. The magnitudes of these currents differ significantly between the models, consistent with the variations seen in Figure \ref{fig:jf}.
Outside the active regions, the models differ in their  distributions of current. The \nlfop{}, \nlfgr{}, \mhs{} and \ffe{} models all show current distributions that correlate with the locations of strongest $|j_r|$ in the photospheric input map (Figure \ref{fig:jr}h). These locations are essentially those with strongest large-scale $B_r$ in the input map (Figure \ref{fig:petrie}). Compared to the other models, \nlfop{} shows a higher background level of current in the quiet Sun, consistent with its values of $J(r)$ and $F(r)$ at this height in Figure \ref{fig:jf}. By contrast, \nlfgr{} shows a much lower level of background current outside of active regions, due to the assumption that $j_r=0$ in those footpoints.  The \mhdcese{} model has small electric currents at low heights, as previously discussed.

The \mf{} and \mhdmas{} models have additional current concentrations outside the locations of strong observed photospheric $|j_r|$. These take the form of concentrated filament channels, seen in Figure \ref{fig:jr} as parallel lines of $|j_r|$, and lying above polarity inversion lines in the photospheric $B_r$. A good example lies between about $60^\circ$ and $125^\circ$ Carrington longitude in the Southern hemisphere in both models. This current concentration is absent from the other models, and is not seen in the observations at the photospheric level. It is a concentration of electric current density in the coronal volume. We will return to this below in Section \ref{sec:filaments}.

\subsection{Plane-of-Sky Magnetic Structure} \label{sec:fl}

\begin{figure}
\includegraphics[width=\textwidth]{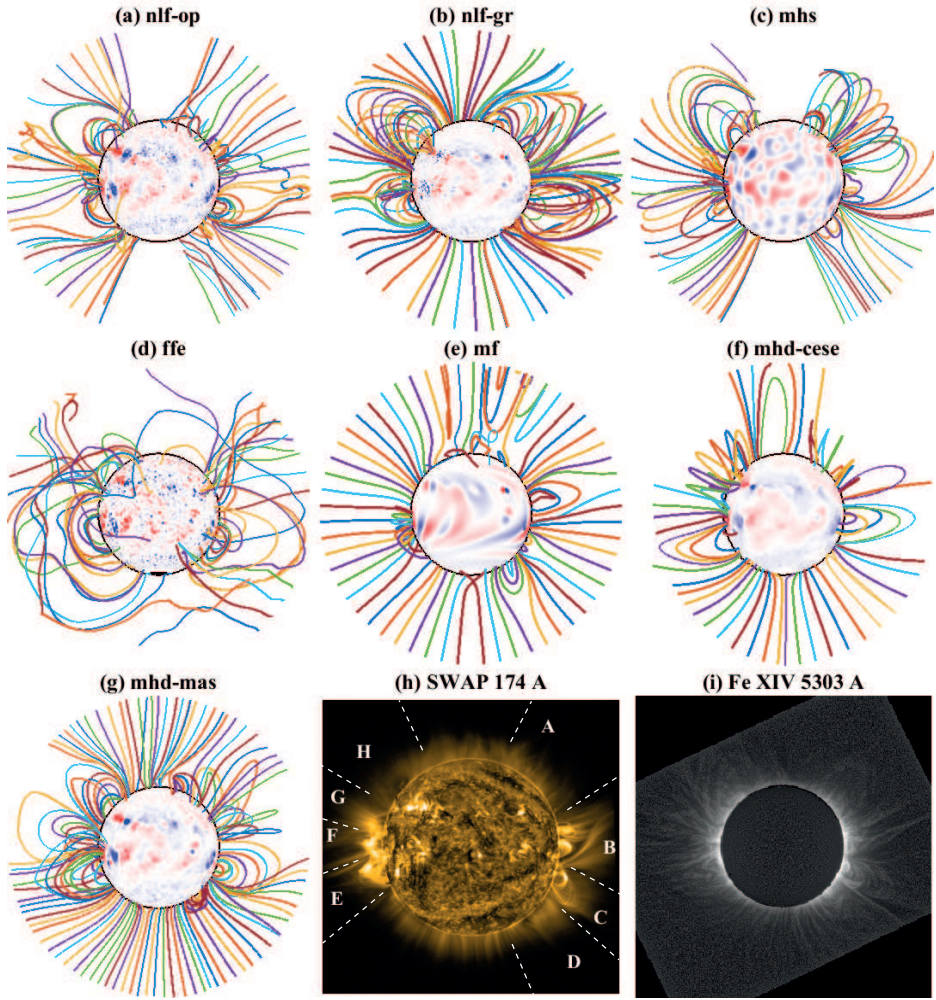}
\caption{Magnetic field structure of each model in the plane-of-sky, as viewed from the Earth at the (approximate) eclipse time. (The density of magnetic field lines is arbitrary and does not correspond to field strength.) Red/blue shading shows $B_r$ on $r=R_\odot$ (saturated at $\pm50\,\mathrm{G}$). Panel (h) shows an EUV 174 \AA{} image of the real corona from PROBA2/SWAP \citep[with median stacking and stray-light correction as described by][]{2013ApJ...777...72S}. Panel (i) shows an image of the corona in the Fe XIV 5305 \AA{} line from the Habbal eclipse expedition, processed \edit{by the MGN algorithm of \citet{2014SoPh..289.2945M} (with parameter $h=0.9$) to bring out the streamer structure.} Several \edit{angular} regions are labelled in panel (h) for ease of reference in the text.}
\label{fig:fl}
\end{figure}

Figures \ref{fig:fl}(a) to (g) show visualizations of each model with  field lines selected to show the plane-of-sky magnetic structure, as viewed from Earth at approximately the eclipse time. \edit{We must bear in mind that, with magnetogram observations presently available only from the Earth's viewpoint, all of the models use primarily synoptic observations built up from central meridian data. These do not include co-temporal information near the limbs, so any comparison can only be approximate. Recall also that the \nlfop{} and \mhs{} models do not include the region poleward of $\pm 70^\circ$ latitude.} The model images can be compared to observations of the real corona in EUV (Figure \ref{fig:fl}h) and the Fe XIV line (Figure \ref{fig:fl}i), as well as in white light out to a larger radius (Figure \ref{fig:druckcomp}). These \edit{particular} observations have been chosen and processed to bring out as clearly as possible the structure of coronal streamers above the limb.
\edit{References describing the processing are given in the figure caption.}
When comparing with the models it must be remembered that the observations see total emission along the line-of-sight, which includes structures in front of or behind the sky plane. It is beyond the scope of this project to do forward modelling of coronal emission, particularly since most of the models are purely magnetic.
Nevertheless, it is clear that there are significant differences between the streamer structure in the different models, as well as similarities.

Firstly, we observe that streamers (closed field regions) are often  too high in many models, particularly \mhdcese{}, \mhs{}, \ffe{} and \nlfgr{}. The white light observations (Figure \ref{fig:druckcomp}) suggest their cusps to lie below $2\,R_\odot$ in most cases, although there could be larger closed loops that cannot be seen due to signal-to-noise issues. In the \nlfgr{} model, these closed-field regions are close to potential, so the cusp height is set by the potential field source surface at $2.5\,R_\odot$. In the  \mhdmas{} model, the outer boundary condition is one of zero velocity, tending also to keep the streamer cusps at the outer boundary. In \ffe{} and \mhdcese{} there is no source surface at all, allowing the closed regions to extend even further out. Similarly, the \mhs{} model is an infinite-space solution with no outer boundary imposed. In the \nlfop{} and \mf{} models the electric currents allow the streamers to have lower cusps. This is further helped in the \mf{} model by the radial outflow which pulls out the field lines. These differences highlight the importance of the outer boundary conditions for this kind of modelling.

Some models show unusual field line behavior near the outer boundary. For example, \mf{} has disconnected U-loops, which are the result of the ejection of a magnetic flux rope \citep{2006ApJ...641..577M, 2014SoPh..289..631Y}. Also, some of the field lines for \nlfop{} and \ffe{} are less smooth near the outer boundary. In the case of \ffe{} this is consistent with the large currents seen at larger heights, and the fact that the relaxation has not reached a stable equilibrium. Although these field lines are strongly non-potential, they contribute little to $\E$ because the magnetic field strength is weak.

We can also compare the angular positions of particular streamers. These are sensitive to both the input data and the locations of electric currents in the corona. For ease of reference, eight approximate locations are labelled in Figure \ref{fig:fl}(h), although these do not necessarily correspond to individual streamers.
Looking first at the West limb (A to D), the most prominent streamers in the observations are at B and C. All models show evidence of closed field near these locations, although their morphology differs significantly between models.
Except for \mhdcese{}, the models also tend to agree that there is a narrow pseudo-streamer structure at D.
\edit{At location A, the observed streamer is less clear in EUV, although there are indications of closed field in Figures \ref{fig:druckcomp} and \ref{fig:fl}(i). Many of the models show at least some closed field at this location, although its structure and orientation are quite varied between models.}
As mentioned above, the \mf{} model shows U-loops here resulting from a recent flux rope ejection. Interestingly, there are indications in SWAP of a high-altitude EUV cavity at this location, although it is clearest on the day before that shown in Figure \ref{fig:fl}(h). Such a cavity is often associated with an underlying filament or filament channel.

On the East limb, the EUV observations indicate prominent structures at E, F and G around the equator, and there is a filament channel at H on the polar crown.
All models indicate the presence of closed field around E, F and G, associated with the several active regions spread around Carrington longitude $0^\circ$.
The \mf{} model has rather less closed field here than the other models, owing to the fact that two of the main bipolar regions have not yet reached the visible face of the Sun and been incorporated into the time-evolving \mf{} simulation. This is an example of when additional magnetogram observations around the East limb, such as might be obtained from a satellite at the L5 Lagrange point, would be beneficial if the simulation were to be carried out in real time \citep{2016ApJ...825..131M}.

At H, all models agree that there is a closed field arcade overlying the East-West polarity inversion line. This is supported by the white light and EUV observations which show a filament at this location.
Sheared magnetic field at low heights corresponding to this filament channel has formed naturally in the \mf{} model, and has correspondingly been added to the \mhdmas{} model. This sheared field was created in \mf{} by differential rotation and flux cancellation at the polarity inversion line, and is not present in the other static models. In these other models, this arcade is closer to potential.

\subsection{Filament Channels} \label{sec:filaments}

\begin{figure}
\includegraphics[width=\textwidth]{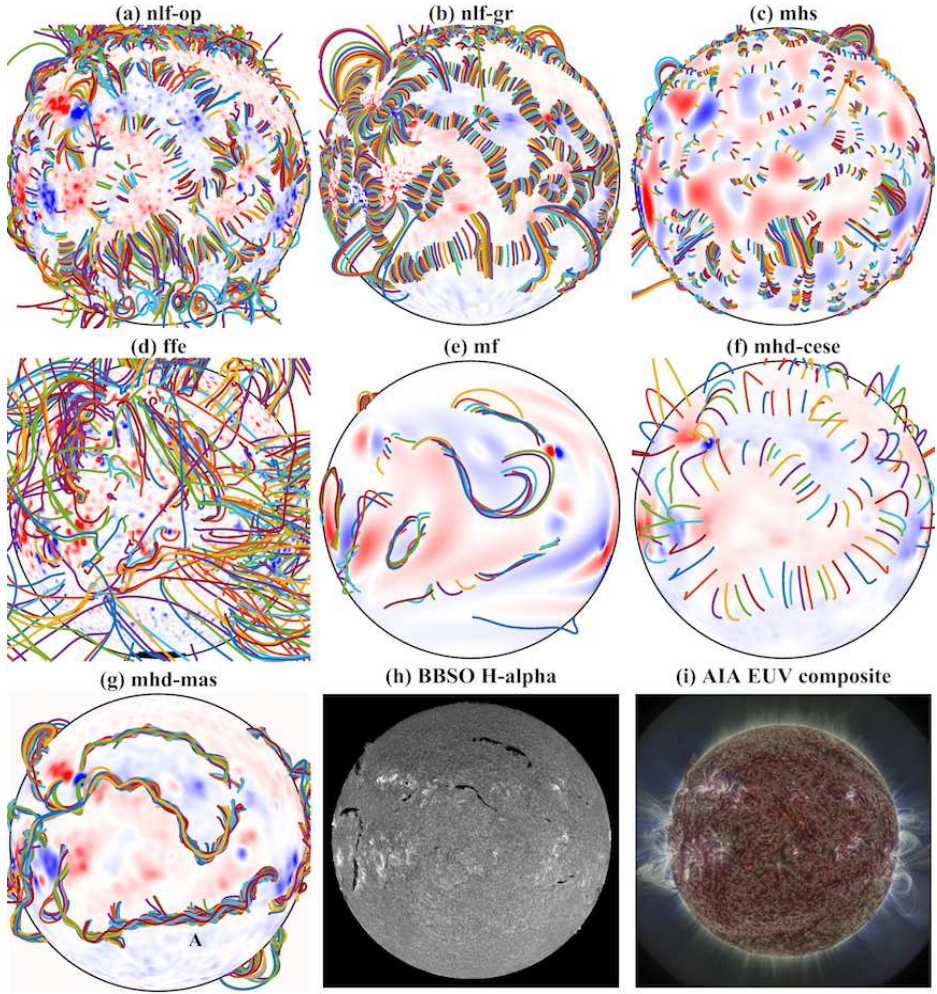}
\caption{Magnetic field lines for each model chosen to show the low coronal magnetic structure above each polarity inversion line on the visible disk. Red/blue shading shows $B_r$ on $r=R_\odot$ (saturated at $\pm50\,\mathrm{G}$). Panel (h) shows an H$\alpha$ image from Big Bear Solar Observatory (taken at 1733UT on the eclipse day). Panel (i) shows a composite of the 171, 193 and 211 \AA{} channels (red, green, blue respectively) from AIA, processed as in Figure \ref{fig:druckcomp}.}
\label{fig:pil}
\end{figure}

An important observable signature of non-potential magnetic structure in the corona is the presence of filament channels and filaments. These are located above polarity inversion lines in the photospheric $B_r$, and are understood to have a strongly sheared magnetic component along the inversion line \citep{2010SSRv..151..333M}. To illustrate this aspect, Figure \ref{fig:pil} shows a selection of magnetic field lines for each model, this time traced from height $r=1.02\,R_\odot$ above polarity inversion lines on the solar disk. Observed images in H$\alpha$ and EUV are shown for comparison.

On the solar disk there are a number of filaments visible in the observations, both in H$\alpha$ (Figure \ref{fig:pil}h) and in EUV (Figure \ref{fig:pil}i). One filament extends around the North-East limb as a prominence in the white-light eclipse image (Figure \ref{fig:druckcomp}). This is the structure labelled H in Figure \ref{fig:fl}(h). The presence of filaments at these particular locations indicates that sheared magnetic field is present.
Only the \mf{} and \mhdmas{} models have significantly sheared magnetic field along polarity inversion lines. In the case of the \mf{} model, this has built up naturally over time due to surface motions and flux cancellation, whereas in the \mhdmas{} model it has been imposed based on the \mf{} model results (Appendix \ref{app:mas}).

In principle, it would be possible for the \nlfop{} and \nlfgr{} models to contain sheared fields along these polarity inversion lines. Indeed the \nlfgr{} model does recover a sigmoidal structure within the active region located around Carrington longitude $190^\circ$ (on the far side of the Sun during the eclipse). However, the transverse magnetic field measurements, from which the currents are determined, have a poor signal-to-noise ratio outside of active regions.  This arises because the linear polarization scales with $|{\bf B}|^2$, in contrast to the circular polarization which is linear in $|{\bf B}|$. Moreover, by creating synoptic maps, these horizontal magnetic fields and resulting currents are further smeared out.
As a result, the corresponding electric currents outside of active regions are not accurate enough to allow the \nlfop{} and \nlfgr{} extrapolations to recover the sheared magnetic fields in filament channels.
In fact, as we have seen, the \nlfgr{} model assumed $j_r=0$ in the weak field regions of the photosphere, owing to this uncertainty.

The resolution of the BBSO H$\alpha$ image in Figure \ref{fig:pil} is sufficient to show filaments themselves, as dark structures, but is not sufficient to show empty filament channels. Thus absence of a filament can not be taken to mean absence of sheared magnetic fields. Filament channels may also be identified from alignment of coronal cells in AIA images, particularly in the 193\AA{}  channel \citep{2012ApJ...749...40S}. Careful inspection of SDO/AIA movies of EUV emission, especially using composite images processed by the \citet{2014SoPh..289.2945M} technique, was used here to identify which filament channels to energize in the \mhdmas{} model (see Appendix \ref{app:mas}). This method identifies many filament channels that do not show up clearly in Figure \ref{fig:pil}(h) and (i). As a case in point, consider the location labelled A in Figure \ref{fig:pil}(g), where a sheared field was inserted in the \mhdmas{} model. There is only a small amount of H$\alpha$ filament material visible at the west end of the channel in Figure \ref{fig:pil}(h), but analysis of the EUV observations at higher resolution suggests the possible presence of a long East-West filament channel all along this polarity inversion line, bending northward at its eastern end as in the \mhdmas{} model. A sheared filament channel is also present here in the \mf{} model. 

\subsection{Open Magnetic Flux} \label{sec:open}

\edit{The open magnetic field lines are the source of the solar wind, so represent the output of the models as far as the heliosphere is concerned. In Section \ref{sec:flux} (Figure \ref{fig:flux}a), we compared the open magnetic flux $\Phi(2.5\,R_\odot)$, as defined in Equation \eqref{eqn:phi}. The numerical values for each model are given explicitly in Table \ref{tab:openflux}, along with an observational estimate based on \textit{in situ} OMNI data. To make this estimate, daily averages of the basic hourly OMNI data were obtained from the GSFC/SPDF OMNIWeb interface at \url{http://omniweb.gsfc.nasa.gov}, giving $B_r(R_{\rm E})$. Averaging over 27 days centered on the eclipse time, and assuming a uniform distribution of magnetic flux over latitude at $R_{\rm E}=1\,\mathrm{AU}$, we estimate the equivalent open flux shown in Table \ref{tab:openflux} as $\Phi(2.5\,R_\odot)\approx 4\pi R_{\rm E}^2|B_r(R_{\rm E})| = 9.05\times 10^{22}\,{\rm Mx}$. As has previously been found during relatively active periods of solar activity \citep[see ][]{2017ApJ...848...70L}, this observed value is much higher than that predicted by models -- either potential or non-potential. Our results are consistent with their findings. The additional coronal currents in the \mf{} and \mhdcese{} models, in particular, enhance the open flux significantly, but it still remains below the level inferred from OMNI observations. The reasons for this discrepancy are not yet understood, and are discussed by \citet{2017ApJ...848...70L}. It is possible that this may be partially heliospheric in origin, if there are regions in the solar wind where the interplanetary magnetic field folds back on itself and is thus over-counted in the $B_r(R_{\rm E})$ measurements \citep{2013JGRA..118.1868O}.}

\begin{table}
\caption{Unsigned open magnetic flux $\Phi(2.5\,R_\odot)$ of the non-potential models.}
\label{tab:openflux}       
\begin{tabular}{lr}
\hline\noalign{\smallskip}
Model & $\Phi(2.5\,R_\odot)\, [10^{22}\,{\rm Mx}]$  \\
\noalign{\smallskip}\hline\noalign{\smallskip}
\nlfop{} & $3.27$\\[0.1cm]
\nlfgr{} & $2.67$\\[0.1cm]
\mhs{} & $4.37$ \\[0.1cm]
\ffe{} & $2.69$ \\[0.1cm]
\mf{} & $5.96$\\[0.1cm]
\mhdcese{} &  $5.30$\\[0.1cm]
\mhdmas{} & $2.97$\\[0.3cm]
OMNI data (27-day average) & $9.05$\\
\noalign{\smallskip}\hline
\end{tabular}
\end{table}

Finally, we consider the \edit{spatial} distribution of open and closed magnetic field in the models. \edit{In addition to indicating the possible source regions of the solar wind, this is important because} the open or closed nature of the magnetic field at a given location depends sensitively on both the input data and the distribution of electric currents in the model. Comparing the open field footpoints to the locations of observed coronal holes therefore provides a further observational constraint.

\begin{figure}
\includegraphics[width=\textwidth]{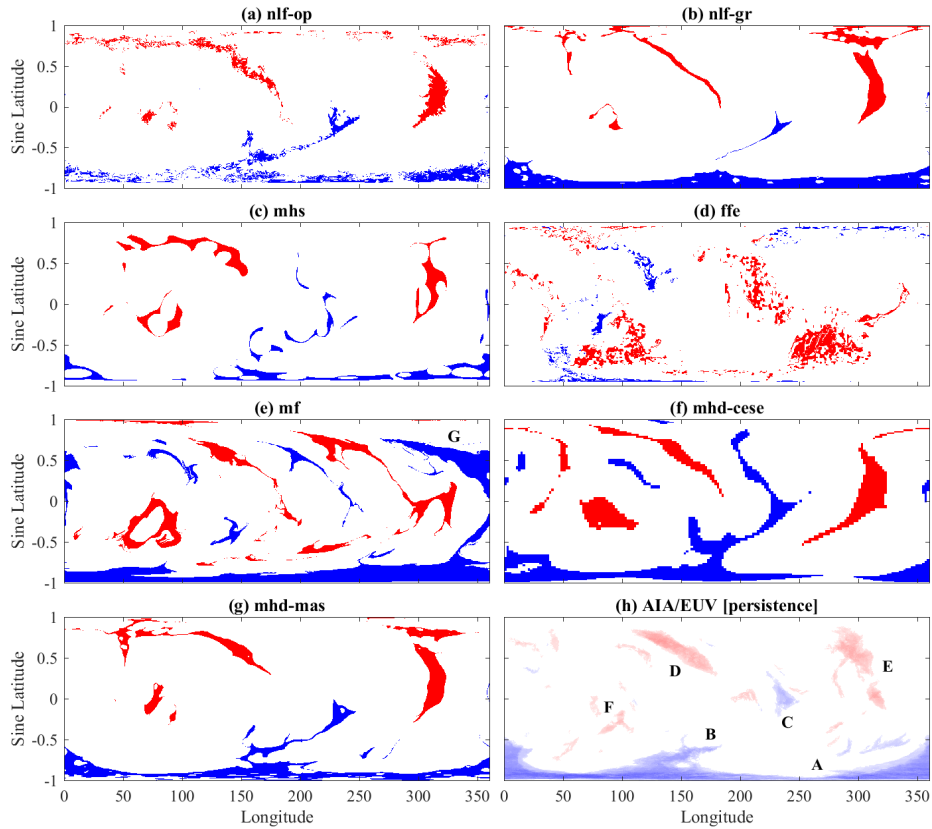}
\caption{Footpoints of open magnetic field lines shown on $r=R_\odot$, and their polarity (red positive, blue negative). Panel (h) shows the observed coronal hole persistence map (see text) using AIA data. In this map, the colors again indicate polarity, with density indicating persistence, i.e., the number of days for which a coronal hole was observed at that location.}
\label{fig:open}
\end{figure}

Figure \ref{fig:open} shows the locations of open magnetic field line footpoints on $r=R_\odot$ for the different models, along with -- in panel (h) -- a persistence map of observed coronal holes in EUV. This map was built up synoptically over Carrington rotation 2161 (2015 February 28 to 2015 March 27) by thresholding full-disk SDO/AIA 193 \AA{} EUV images, at a cadence of 12 hours. The procedure is described in more detail by \citet{2017SoPh..292...18L}, although during this particular time period no far-side EUV data were available. Of course, this should be taken only as a lower bound on the areas of observed open field rather than a true measure. This is because, while a significant fraction of open flux originates from coronal holes, the footpoints of open field can also be bright, particularly if they lie in active regions. \edit{Indeed, quantifying the amount of open flux not located in coronal holes is an important observational challenge for reconciling the models and observations in future.}

Given the sensitivity of the open/closed footpoint regions as well as the variety of boundary conditions used, there is reasonable agreement between most of the models and with the observed persistence map. Robust features are labelled in Figure \ref{fig:open}(h) and include: a negative-polarity polar coronal hole in the Southern Hemisphere (A) but no corresponding hole in the Northern hemisphere; a narrow equatorward extension of the Southern polar hole (B and C); a long positive-polarity hole in the Northern hemisphere  (D); a north-south oriented positive-polarity hole (E); and a more compact positive-polarity hole near the equator in the Southern Hemisphere (F).

The \ffe{} model differs significantly from the others, as was evident from its magnetic field structure in Figure \ref{fig:fl}.
The \mf{} model reproduces many of the observed coronal holes, although some (particularly E) are shifted in position. However, it has in a number of additional open field regions, with significantly more open magnetic field of both polarities at low and medium latitudes. Indeed this additional open flux was seen in Figure \ref{fig:flux}(a). Partly this additional open field arises from the opening up of streamers discussed in Section \ref{sec:fl} and seen in Figure \ref{fig:fl}. Another reason for enhancement is the ejection of magnetic flux ropes in the \mf{} model (one such eruption is responsible for the U-loops in Figure \ref{fig:fl}e). But the difference from other models is also partly due to the different photospheric distribution of $B_r$ (Figure \ref{fig:br1}), owing to the model being driven by a flux transport model rather than directly from observed magnetograms (Appendix \ref{app:mf}). A good example is the large extension of the negative-polarity hole into the Northern hemisphere around Carrington longitude $300^\circ$ to $360^\circ$ (labelled G in Figure \ref{fig:open}e). This hole is likely to reduce in size once the new active regions emerge that are already present in the other models.

\begin{figure}
\includegraphics[width=\textwidth]{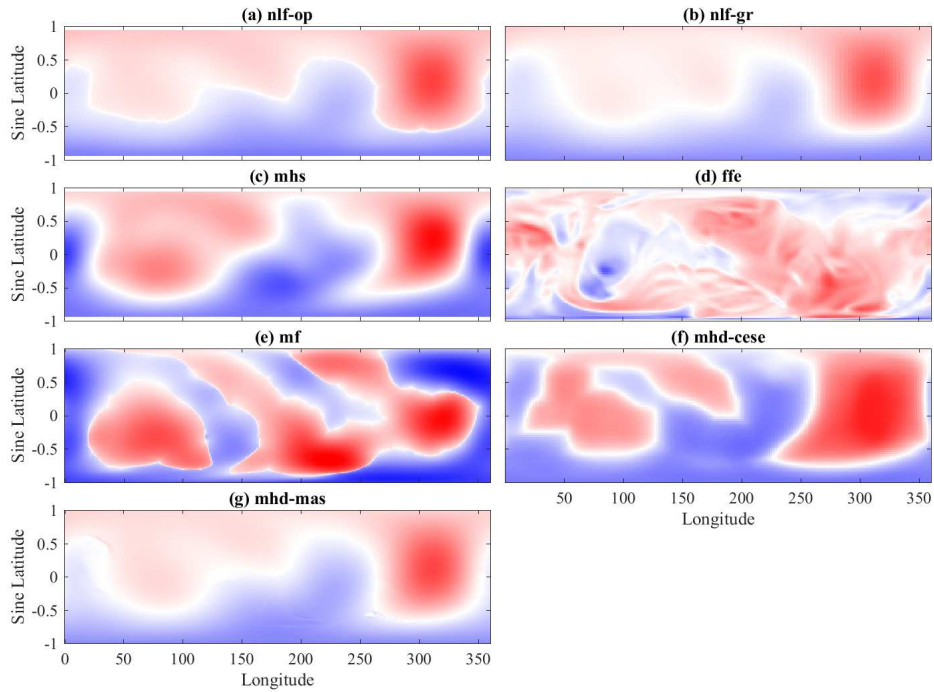}
\caption{Radial magnetic field at $r=2.5R_\odot$ in each model, over the full spherical surface. The color tables are saturated at $\pm 0.4\,\mathrm{G}$.}
\label{fig:br2_5}
\end{figure}

Figure \ref{fig:br2_5} shows the distribution of $B_r$ on the outer boundary $r=2.5\,R_\odot$ in each model. Here all of the field lines are ``open'' according to our definition.
At this radius, the magnetic field is insensitive to small-scale differences in the input magnetic maps, and there is reasonable agreement between \nlfop{}, \nlfgr{}, \mhs{} and \mhdmas{}. These models all inherit the same basic open/closed topology from the potential field used as their initial condition before the injection of electric currents at the base.
In the \ffe{}, \mf{} and \mhdcese{} models, the heliospheric current sheet (boundary between positive and negative $B_r$) has a more complex shape, with a disconnected loop in the \mf{} model. The reasons for these differences are the same as in Figure \ref{fig:open}. As mentioned above, the \ffe{} model has a complex magnetic structure that likely reflects the incomplete relaxation of the model. These results do suggest that we should not take the potential field topology for granted, although some of these differences arise from the different  photospheric $B_r$ distribution in the \mf{} and \mhdcese{} models; indeed, the coronal hole map comparison suggests that the difference may be over-emphasized in these models.
The additional open flux in the \mhs{}, \mhs{} and \mhdcese{} models, compared to the others, is also clear in this plot. Note that this could be reduced in the \mhs{} model by reducing the parameter $a$, so may not be particularly significant in that case.

\section{Conclusions} \label{sec:discuss}

Having analysed seven different non-potential models of the coronal magnetic field on 2015 March 20, we now draw some overall conclusions. 

The initial impression from Figure \ref{fig:fl}, for example, is one of significant disagreement between different models. This is true particularly in regard to the overall magnetic structure.
This disagreement arises from several sources: the input data used by each modeller, the coronal modelling techniques themselves, and the outer boundary conditions.

There are two fundamental limitations with currently available magnetogram input, not including the differences between magnetograms from different instruments and observatories \citep{2014SoPh..289..769R}, which we avoid here by focusing on SDO/HMI. One limitation is the lack of co-temporal full surface coverage, which necessitates the use of synoptic maps that combine observations taken at different times. 
The different models achieve this in different ways, as described in Section \ref{sec:models}, and this leads to different maps of the magnetic field across the whole solar surface (Figure \ref{fig:br1}), with corresponding differences in coronal magnetic topology. 
One consequence of the use of synoptic maps is that the majority of models (except for \mf{} and \mhdcese{}) use data taken after the eclipse time, meaning that this exercise is not equivalent to a prediction of the eclipse magnetic field which could have been made in advance.
The $B_r$ distribution in the \mf{} model differs particularly from the others because it uses  a surface flux transport model to evolve the photospheric magnetic field over a period of months, rather than inserting magnetogram data more directly. This has the advantage of allowing the build-up of free magnetic energy in weak-field regions, but does mean that there are resulting differences in the magnetic structure.
\edit{The use of synoptic data also means that the magnetogram input near the limbs -- particularly the East limb -- is out-of-date, and this can particularly affect comparisons with eclipse images. The proposed space mission to the L5 Lagrange point would greatly improve the situation at the East limb, provided that a magnetograph were included on board.}

The second limitation is with vector magnetograms. Firstly, the signal-to-noise ratio remains low in the transverse components of ${\bf B}$, so that the \nlfop{}, \nlfgr{} and \ffe{} models do not reproduce the sheared magnetic fields in filament channels.
And the fact that the photospheric vector magnetograms do not satisfy ${\bf j}\times{\bf B}=0$ leads to problems in particular with the \ffe{} model, preventing it from reaching equilibrium. Further refinement will be needed before this model can be of practical use for the solar corona.
In future, it is hoped that this issue will improve if and when upper-chromospheric magnetogram observations become available.

In the coronal volume itself, we have shown that the different models vary in their degree of non-potentiality, as measured by either electric currents or free energy. The models with greater free energy achieve this with a variety of different current distributions: currents may be concentrated in active regions (\nlfgr{}), in filament channels (\mhdmas{} and \mf{}), or may be more distributed throughout the corona (\mf{} and \mhs{}). This aspect of the models is perhaps the most difficult to calibrate against observations, since no direct observations of coronal electric currents are available. The presence of filament channels, for example, illustrates the importance of the gradual build-up of coronal electric currents over time. 
Although these are not reproduced in the \nlfgr{}, \nlfop{}, \mhs{} and \mhdcese{} models, this is not due to a fundamental limitation of the equations used to model the magnetic field in the corona, but due to limitations in the available boundary data.
In particular, we have shown with the \mhdmas{} model how static models can be further energized by the injection of electric currents in filament channels, and this could be applied in future with the other models.
This hybrid approach of informing static models with the results either from simplified time-evolving models (like \mf{}) or from additional coronal observations could well be a useful one in the absence of improved magnetogram input data.

The outer boundary conditions also have a significant impact, and deserve greater attention. With the exception of \mhdcese{}, none of the models presented here couples physically to a solar wind solution beyond $2.5\,R_\odot$. This is because the increasing plasma-$\beta$ in that region no longer allows for the use of a purely magnetic model. Several of the models impose an artificial source surface at $2.5\,R_\odot$, as in the common PFSS model, and this leads to inaccuracies in the streamer structure, and potentially also the open flux. The \mf{} model instead uses a radial outflow boundary condition to mimic the effect of the solar wind, but this seems to be inflating the field too much in this particular case, at least in terms of open field footpoints (though interestingly not in the height of the streamers). The corresponding currents higher in the corona change the open-closed magnetic topology significantly. Overall, this outer boundary is an important problem that requires more sophisticated MHD modelling that includes plasma thermodynamics. Such simulations have been performed with the full-MHD version of the \mhdmas{} model \citep[e.g.,][]{2009ApJ...690..902L, 2013Sci...340.1196D}, which can describe the solar wind.

All of this being said, there are also areas of broad agreement between many of the models. For example, while they have very different input grid resolutions on $r=R_\odot$, this \edit{(in itself)} does not affect the estimated open flux and topology of the heliospheric current sheet on $r=2.5\,R_\odot$, \edit{which arises rather from the differences between the coronal modelling approaches}. On $r=R_\odot$, the footpoint regions of open magnetic field show similarities in all models, and there is agreement that the strongest currents are within the active regions. Among those models with significant free magnetic energy, there is general agreement on the ratio $\E/\Ep\approx 1.4$ to $1.5$. And the locations of closed field streamers are broadly in agreement, though not their height and shape.

From this study it is clear that all of the models have positive aspects that agree with observations, but other aspects that do not match so well. Much of this can be related to the distribution of electric current both within and outside of active regions. 
At present, nonlinear force-free extrapolations such as \nlfop{} or \nlfgr{} are best at representing the structure of active regions where reliable vector magnetic field input data are available.
But accounting for the free energy outside of active regions is currently possible only with time-evolving models such as \mf{}.
Yet, in these models, it is  too computationally expensive to account for the full plasma thermodynamics, something that has not been considered here but is already possible in state-of-the-art full-MHD models, albeit for static configurations.

What is clear is that this is an exciting time for coronal magnetic field modelling, with progress on several fronts but much still to do.
Rather than simply waiting for better magnetogram data, our comparisons with currently available observations -- though qualitative -- do suggest that these indirect observational constraints could be better used to optimize the models. The challenge is to do this systematically. The ideal model would match EUV observations of filament channels and coronal loops, the positions of white-light streamers, and the locations of observed coronal holes. A more sophisticated approach would involve forward modelling of actual observed emission, but we suggest that much can already be learned from morphological comparisons. 

\begin{acknowledgements}
The authors gratefully thank ISSI Bern for hosting our International Team and enabling this study to be carried out. The \nlfgr{} work was granted access to the HPC resources of CINES/IDRIS under the allocation 2016-16050438 made by GENCI. ARY and DHM thank the UK STFC for financial support. ARY and CAL were supported by a grant from the US Air Force Office for Scientific Research. DHM also thanks the Leverhulme Trust for financial support. GP acknowledges support from NASA grant NNX15AN43G. FXS was supported by the National Natural Science Foundation of China (Grant No. 41531073). TW acknowledges DFG-grant WI 3211/5-1. Authors from Predictive Science, Inc. were supported by NASA, NSF, and AFOSR, including support from NASA grant NNX15AB65G.  Their computational resources were provided by NASA's Advanced Supercomputing Division and NSF's XSEDE program. LAU was supported by the National Science Foundation Atmospheric and Geospace Sciences Postdoctoral Research Fellowship Program. National Center for Atmospheric Research is sponsored by the National Science Foundation. HM's research is partly funded by the Leverhulme Foundation and an STFC PR\&D grant. The eclipse data is provided through the kind permission of Prof Shadia Habbal at the University of Hawaii who leads an international team of eclipse scientists.
\end{acknowledgements}

\bibliographystyle{spr-mp-sola}       

\appendix
\section{Details of the \mhdmas{} model} \label{app:mas}

The MAS time-dependent magnetohydrodynamic (MHD) model \citep{1994ApJ...430..898M,1995ApJ...438L..45L,1999PhPl....6.2217M} of the global solar corona has been in development for over two decades.  The simulation \mhdmas{} illustrated in this paper uses a simplified zero-beta version of the model, in which pressure and gravity forces are neglected, in order to emphasize the non-potential aspects of the corona, in particular the sheared fields expected in filament channels. An energization mechanism was applied to create sheared and twisted magnetic fields and flux ropes in the observed filament channels.  Unlike the more sophisticated version of the MHD model, this zero-beta version  is not able to include explicitly the effect of the solar wind, but in the future it is hoped to include this energization capability into the full MHD model.

The filaments (visible for example in Figure \ref{fig:pil}h) are thought to be supported by twisted flux-rope-like magnetic fields.  These were introduced in an \textit{ad hoc} manner as follows:
\begin{enumerate}
\item Start with a PFSS solution but add additional flux to $B_r$ on the boundary $r=R_\odot$, to compensate for that lost by the cancellation in Step 3.
\item Emerge transverse magnetic field by imposing a transverse electric field ${\bf E}_t = \nabla_t\Phi$ at the boundary $r=R_\odot$. Here $\Phi$ is chosen to be localized to the filament channels, and reverses sign across the PILs, and $\nabla_t$ is the transverse gradient. Note that this does not change $B_r$.
\item After the transverse magnetic field emergence is complete, cancel flux at the PILs by applying a transverse electric field of the form ${\bf E}_t=\nabla_t\times\psi\hat{\bf r}$, so as to form more flux-rope like field lines out of highly sheared arcades. The function $\psi$ is driven by the change in $B_r$,
$$\frac{1}{c}\frac{\partial B_r}{\partial t}=
\mathbf{\nabla\!}_t^{\,2}\psi{\hskip 2pt}{\rm.}$$
By the end of this process, $B_r$ will once again match the magnetogram, thanks to the additional flux added in Step 1.
\end{enumerate}
 The total amount of added flux, which was chosen by trial and error, corresponds to 20\% of the flux in the magnetogram (over the whole Sun).  Since the added flux is localized to filament channels, it corresponds to a greater fraction of the flux in the filament channels (on the order of $40\%$).

\begin{figure}
\begin{center}
\includegraphics[width=\textwidth]{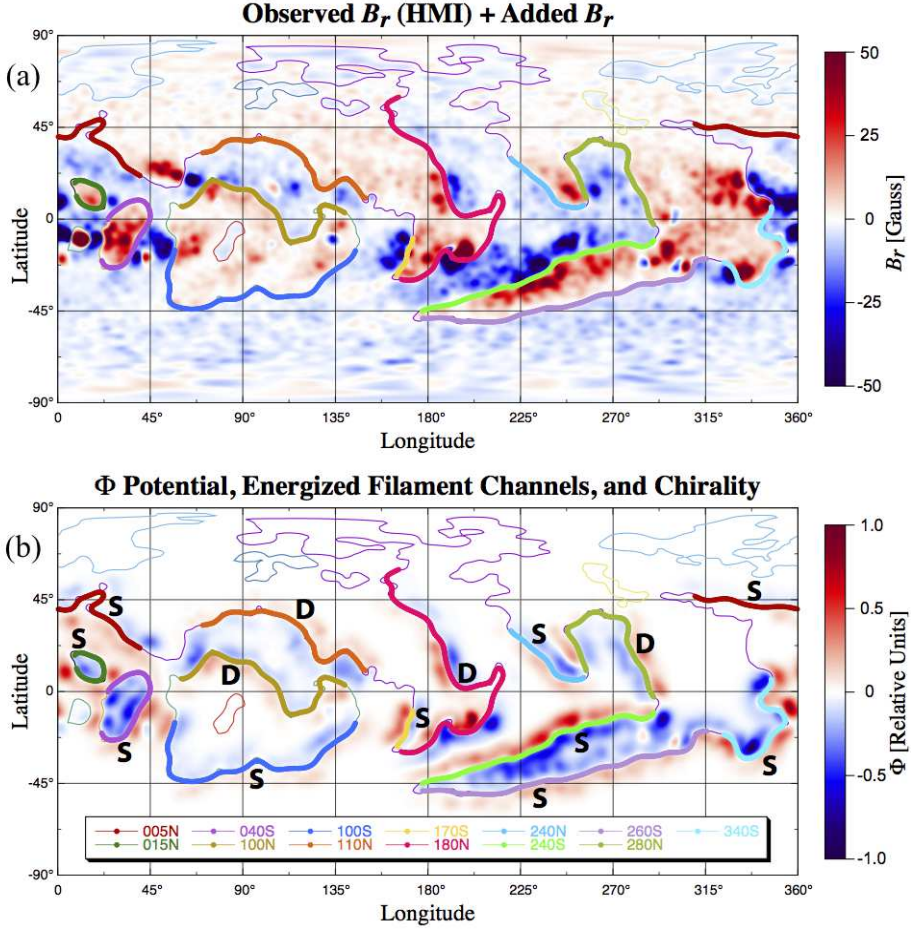}
\end{center}
\caption{Energization of filament channels in the \mhdmas{} model. Panel (a) shows radial magnetic field $B_r$ at $r=R_\odot$ for the observed HMI field with the added flux.  The colored contours are the PILs in the PFSS model at $r=1.02R_\odot$, with thick colored line segments showing the filament channels in which the field was energized. Panel (b) shows the energizing $\Phi$ potential.  The chirality selected for the magnetic field in these filament channels is shown in (b), with ``S'' denoting sinistral and ``D'' denoting dextral filaments.}
\label{fig-ZM1}
\end{figure}

The filament channel regions at which the field was twisted in this manner were chosen in an {\it ad hoc} manner, guided by both the results of the \mf{} model and the locations of filament channels inferred from SDO/AIA EUV movies during the evolution surrounding the eclipse.  Since the \mf{} model assimilated the emerging active regions from HMI observations over time, it was able to give the best estimate of the chirality of the fields in the filament channels.  By chirality here we mean the direction of the transverse field introduced.  According to the usual definition, a sinistral filament has an axial field that points to the left when looking towards the PIL from the side with $B_r>0$, and vice versa for a dextral filament.  In the technique described above, choosing $\Phi$ with the same sign as $B_r$ produces a dextral filament, whereas $\Phi$ with opposite sign to $B_r$ produces a sinistral filament.  In some cases, when the chirality changed within a single filament,  a single chirality was used to simplify matters.  Figure~\ref{fig-ZM1} shows the filament channels where the magnetic fields were energized, with the chosen chirality (as deduced from the \mf{} model), and the corresponding $\Phi$ potential used. The magnetic energy of the initial PFSS field (with the observed $B_r$ plus the added flux) was $2.13\times 10^{33}\,{\rm ergs}$.  The energy in the PFSS field corresponding to the observed $B_r$ was $1.72\times 10^{33}\,{\rm ergs}$.  The energy in the final field, with twisted fields in the filament channels, matching the observed $B_r$, was $2.52\times 10^{33}\,{\rm ergs}$, so that the final field had an energy $47\%$ above the corresponding PFSS field.

\section{Details of the \mf{} model} \label{app:mf}

For the \mf{} calculation presented here, the coronal magnetic field was evolved for 200 days up to the eclipse date, allowing the self consistent build-up of electric currents and free magnetic energy, driven by the photospheric evolution (both footpoint motions and flux emergence). The simulation was initialized with a potential field extrapolation for 2014 September 1, and a fully non-potential corona was arrived at after 6-8 weeks (as measured by average current density), well before the eclipse date.

Compared to previously published simulations, a different evolution of $B_r$ at the solar surface $r=R_\odot$ was used. Namely, this was derived from the Advective Flux Transport (AFT) model of \citet{2014ApJ...780....5U,2014ApJ...792..142U} and \citet{2015ApJ...815...90U}, which assimilates HMI magnetograms on the visible disk and advects the distribution over the full solar surface using differential rotation, meridional flow and small-scale convective flows. For driving the magneto-frictional model, it is actually the horizontal electric field that is required. Since this is unknown in the regions assimilated from HMI, the AFT model could not be used directly to drive the 3D simulations. Instead, it was used to determine the locations of new bipolar magnetic regions (BMRs), which were then inserted into a simpler flux transport model which also includes differential rotation and meridional circulation, but approximates the small-scale convective flows by a simple supergranular diffusion term. The resulting smoother distribution of the photospheric field is evident in Figure \ref{fig:br1}(e).

The BMR properties were derived from the AFT model by a three-stage automated procedure:
\begin{enumerate}
\item Compare successive $B_r$ maps (once every 24 hrs) to identify new BMRs (leading to 197 distinct BMRs).
\item Compute the BMR properties (location, size, magnetic flux and tilt angle) once per hour, from 48 hrs before to 120 hrs after initial identification, and determine the time of maximum flux. Insert the (3D) BMR in the new simulation at this time with the corresponding properties.
\item Check the resulting $B_r$ evolution manually, and correct the BMR properties to best reproduce their times of emergence and also the structure of multiple-BMR activity complexes.
\end{enumerate}
In the run shown here, each BMR was given a twist when inserted into the 3D simulation -- so as to model the active region helicity -- with magnitude $\beta=\pm0.4$ \citep[see][]{2008SoPh..247..103Y}.

\end{document}